\documentclass[11pt, onecolumn]{IEEEtran}

\usepackage{graphics} 
\usepackage{epsfig} 
\usepackage{mathptmx} 
\usepackage{times} 
\usepackage{amsmath} 
\usepackage{amssymb}  
\usepackage{cite}
\usepackage{flushend}
\usepackage{epstopdf}
\usepackage{graphicx}
\usepackage{caption}
\usepackage{subcaption}

\newcommand{\ma}{\left[ \begin{array}}
\newcommand{\ema}{\end{array}\right]}
\def\beq{\begin{equation}}
\def\eeq{\end{equation}}
\def\beqa{\begin{eqnarray}}
\def\eeqa{\end{eqnarray}}
\def\ba{\begin{eqnarray*}}
\def\ea{\end{eqnarray*}}
\def\bc{\begin{center}}
\def\ec{\end{center}}

\def\bcase{\left\{ \begin{array}{ll}} 
\def\ecase{\end{array} \right.}

\vspace{3mm}
\title{\LARGE \bf Continuous Circadian Phase Estimation Using Adaptive Notch Filter}

\author{Wei Qiao$^*$, Kyle Altman$^*$, Agung Julius$^*$, Bernard Possidente$^\dagger$, John T. Wen$^*$, \\ 
$^*$ Smart Lighting Engineering Research Center, Rensselaer Polytechnic Institute, Troy, NY\\
$^\dagger$ Biology Department, Skidmore College, Saratoga, NY}

\begin{document}
\bibliographystyle{unsrt}

\maketitle
\thispagestyle{empty}
\pagestyle{empty}
\newtheorem{remark}{Remark}


\begin{abstract}
Actigraphy has been widely used for the analysis of circadian rhythm.
Current practice applies regression analysis to data from multiple days to estimate the 
circadian phase.  This paper presents a filtering method for online processing  of 
biometric data to estimate the circadian phase. We apply the proposed method on actigraphy data
of fruit flies (\textit{Drosophila melanogaster}).
\end{abstract}

\section{Introduction}\label{intro:sec}
Circadian rhythms are unique features of life that terrestrial species have evolved in response to the 24 h cyclic environment of the Earth's surface, which reflects the natural light cycle. For example, in humans, molecular oscillations of the biological clock drive rhythmic physiological and behavioral functions i.e., heart beat, blood pressure, body temperature, sleep-wake cycle, metabolism and locomotor activity. The disruption of circadian rhythm caused by lack of synchrony between the circadian clock in the brain and the external environment may produce serious detriments in human health and well-being and result in various issues ranging from increased sleepiness, decreased attention span and lower productivity, to long-term health problems such as increased risk for cancer, diabetes, obesity, and cardiovascular disorders \cite{kripke1978circadian,knutsson03,sephton2003circadian,stevens2005circadian,Rea08}.
Such disruption may be often observed in people with irregular sleep patterns, artificial deprivation of light, i.e., submariners or mine workers \cite{kelly1999nonentrained,mills1964circadian}, frequently shifted sleep-wake cycles of night nurses \cite{knutsson03}, and travelers who cross multiple time zones \cite{harrington10}. Therefore, control of circadian rhythm is of great importance to circadian research.

Among many organisms, \emph{Drosophila} has been widely studied in the investigation of circadian rhythms since the 1950's starting with the frontier work of C. Pittendrigh. Pittendrigh's studies mostly focused on the eclosion rhythms of Drosophila: the emergence of larva from the pupa case; in recent decades, genetic analysis based bio-molecular interaction studies attracted many researchers to investigate circadian rhythm of \emph{Drosophila} experimentally and led to unveiling the molecular interactions of PER, TIM proteins and its mRNA, based on which \emph{Drosophila} circadian models are proposed \cite{Leloup99}. However, the high cost and complexity of these protocols prohibit the continuous measurement of circadian rhythm outputs; thus another behavioral genotype, the locomotor activity, is often chosen as the output of the circadian clock. Although not as accurate as genetic analysis, the advantages of locomotor activity based circadian research are low-cost, continuous measurement and integration as a fully automated control and measuring system. Different types of \emph{Drosophila} activity recording system were developed in the past decades; the most widely used method is the commercial TriKinetics (www.trikinetics.com), which enables a fully automated recording system.

The traditional method of circadian phase estimation is mostly based on double-actogram, in which phase is extracted by Onset time, Acrophase (peak of a fitted cosine curve) or an eye-fit line. Despite the inaccurate nature of this method, the phase estimation is on a daily basis and does not provide continuous phase estimation, which would be an obstacle in many practices other than obtaining phase response curve, especially in feedback control problems, which relies on circadian phase, i.e., a jet-lag problem.

In this paper, we adopt a model-free circadian phase estimation scheme using a modified adaptive notch filter (ANF) developed by our group previously \cite{ZhangACC13} (inspired by \cite{Hsu1999}), and design a set of experiments to obtain the phase response curve. With ANF, we can continuously estimate circadian phase based on the locomotor activity of \emph{Drosophila}. This tool, which enables continuous circadian phase estimation, will be of vital importance to any circadian related problems, is especially a big step towards closed loop circadian rhythm control in \emph{Drosophila} and is a helpful tool for a more complex study of human circadian rhythm. We will show in this paper that the ANF method not only estimates the circadian phase accurately and continuously, but also offers filtering effects that can successfully estimate the circadian phase from a set of noise-corrupted data, which the traditional double-actogram method fails in.

The paper is organized as follows: some preliminary results of circadian phase estimation with ANF are introduced in Section \ref{Pre:Sec}. The experiment setup and protocols are detailed presented in Section \ref{Protocol:Sec}. The experiment design in obtaining PRC is introduced in Section \ref{Exp:Sec} followed by the results and discussion in Section \ref{Result:Sec}. The concluding remarks are given in Section \ref{Conc:Sec}.

\section{Preliminaries}\label{Pre:Sec}
In this section, how \emph{Drosophila} locomotor activity is affected by external light is discussed and some preliminary results of circadian phase estimation are introduced.
\subsection{External Light and Drosophila Locomotor Activity}
Similar to that of humans, \emph{Drosophila}'s circadian system is most sensitive to blue light and almost completely insensitive to red light \cite{Frank69,Klarsfeld03}. \textbf{CRY} (cryptochromes) in \emph{Drosophila}'s deep brain is the circadian photorecepter, which regulates the impact of light on the circadian system. The circadian pacemaker, affected by CRY, regulates the expression of neuropeptide PDF, which affects the locomotor activity \cite{Emery00}. Therefore, \emph{Drosophila} locomotor activity can be used as a prediction of its circadian phase.

However, activity is not solely controlled by its circadian pacemaker, but also affected by the visual pathway: light perception by the compound eye and other eye structures of \emph{Drosophila}. It can be observed in the experiments that the locomotor activity peaks right after the light turns on or off even for red light.

to be continued...

\subsection{Adaptive Notch Filter and Locomotor Activity based Circadian Rhythm Estimation}\label{ANF:Sec}
Measuring circadian phase experimentally in real time is challenging; the most reliable method is genetic analysis for \emph{Drosophila} and onset of melatonin secretion under dim light conditions (dim light melatonin onset, or DLMO) for human through lab test on saliva or plasma samples. However such measurements are inconvenient, time consuming, and expensive. Some circadian-related physiological signals such as activity, heart rate, and body surface temperature, which are possible to be measured at high sampling rate and show rhythmic patterns, are more suitable for real time circadian phase estimation. Therefore, in this paper we focus on \emph{Drosophila} locomotor activity as the circadian clock output. Although simple and easy to measure, such signal is usually noisy and corrupted; therefore a filter is required to extract circadian phase from such noisy physiological measurements.

\begin{figure}[ht]
\bc
  \includegraphics[width=0.45\textwidth]{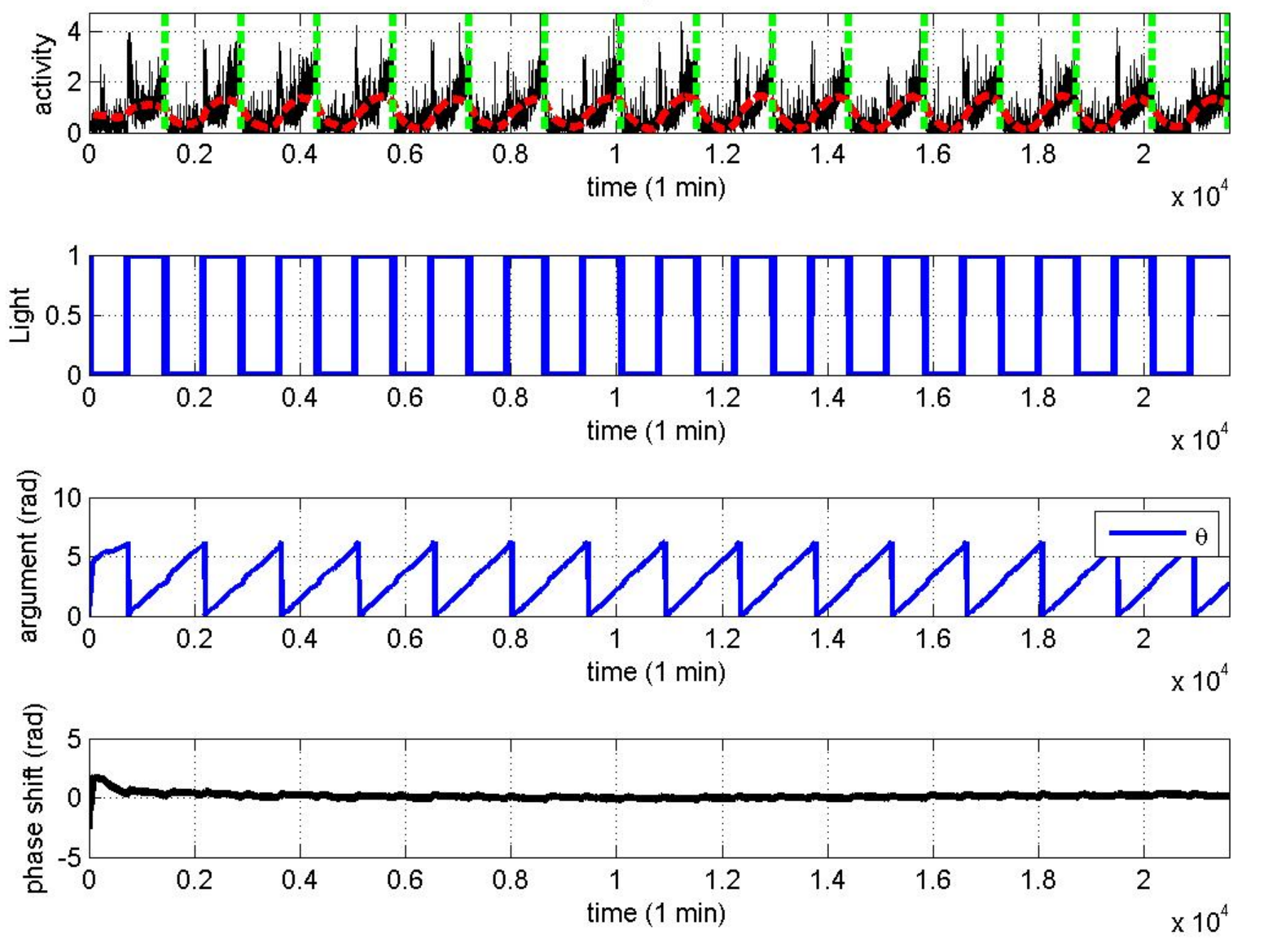}
\ec
  \caption{An example of ANF algorithm on estimating the circadian phase of \emph{Drosophila} activity data. Top: {\em Drosophila} activity (black) and ANF estimation (red). Middle (top): Periodic blue light. Middle (bottom): the ANF extracts the estimated signal argument. Bottom: the estimated phase is obtained by subtracting the linearly increasing part from the argument.}\label{free_run:fig}
\end{figure}

A model-free circadian phase estimation scheme using a modified adaptive notch filter (ANF) developed by our group previously \cite{ZhangACC13} (inspired by \cite{Hsu1999}), which can also accommodate the non-zero mean and non-sinusoidal waveform of the circadian signal, is introduced as follows:

Assume the following form of the circadian signal $y$,
\begin{equation*}
   y(t)= \sum_{k=1}^N a_k \sin(k\omega^* t + \phi_k) + d + w(t)
\end{equation*}where $d$ is a constant bias, $w$ is a zero-mean white noise. The proposed modified ANF is given by

\begin{eqnarray}
  \dot{x} &=& A_{ANF}(\omega)x+B_{ANF}(\omega)y,\label{ANF1}\\
  \dot{\omega} &=& \gamma_\omega f_{ANF}(y,\omega, x).\label{ANF2}
\end{eqnarray}where the parameters and tuning procedures can be found in \cite{ZhangACC13}. The ANF produces an estimate for ($\omega^*t+\phi_1$), which is the argument of the fundamental harmonic and is used as the circadian phase estimate. In the cited work, the local stability and robustness properties of the modified ANF algorithm are established and its effectiveness is demonstrated on both synthesized periodic signals and the {\em Drosophila} activity data (an example is shown in Figure \ref{free_run:fig}).

\section{Experiment Setup and Protocol}\label{Protocol:Sec}
\textbf{MATERIALS:}
Drosophila Culture Kits from the Carolina Biological Supply Company are used for breeding and preparation of the experiments. Each kit includes 36 culture vials in which the flies are bred and stored, instant \textit{Drosophila} medium food that sustains the flies in the culture vials, and sorting brushes. A FlyNap Anesthetic Kit, also from Carolina, is used for sedation of the flies. The kit includes a 100 mL vial of FlyNap anesthetic and several anesthetic transfer containers.

\textbf{Chemicals and their application:}
The Flynap anesthetic is a mixture of 50\% triethylamine, 1.25\% 2-propanol, 1.13\% methanol, 25\% neutralizer fragrance, and 22.63\% ethanol. 3 mL of FlyNap poured into the transfer container is enough to anesthetize the flies for at least 30 minutes.

\textbf{Flies:} Canton S wild type. All flies selected for inclusion in the experiment had undergone eclosion three days before the experiment began. 

\medskip

\textbf{EXPERIMENTAL PREPARATION:}
Experimental set-up: 189 flies (21 flies in each incubator) were placed in tubes (one tube per fly) that contained sufficient food for the experiments duration on one side: this food was standard agar food that did not contain pymetrozine. These tubes were then loaded into nine separate Trikinetics Drosophila Activity Monitor (DAM) 5 activity monitors. The temperature is kept at 25 degree Celsius using a temperature control system.

\medskip
\textbf{EQUIPMENT:}

\begin{itemize}
\item[(1)] Trikinetics Drosophila Activity Monitor (DAM) 5: The monitors shine an infra-red beam down the center of each tube and count the number of times the beam is broken by the fly in order to calculate how many times each fly move along the tube. The monitors sum the count data into 1 minute bins and send the data to a PC.\\
\item[(2)] Culturing Incubators: The culture of flies are placed in light-tight incubators with Philips Rebel LEDs of cool white light under 12 h light 12 h dark (LD 12:12) cycles. \\
\item[(3)] Experiment Incubators: The DAM 5 activity monitors are placed in light-tight incubators with Philips Rebel LEDs at different wavelengths as light sources and mirrors mounted on the inner surface of the incubators. The positions of LEDs are optimized by the optical simulator, ZEMAX, to ensure the light uniformity on the monitor.\\
\item[(4)] Temperature Control System: High pressure air through a humidifier is used for ventilation and keeping the food from drying. Temperature sensors and heaters are mounted on each incubator to maintain the temperature in each incubator at 25 degree Celsius.\\
\item[(5)] Light control System: LEDs at different wavelengths mounted on top of each incubators are controlled by a PC with designated light patterns and controllable light intensity.
\end{itemize}

\medskip
\textbf{Software:}
We use MATLAB for most of data handling and analysis and implementation of our algorithms. ActogramJ \cite{ActoJ} is used to obtain the Acrophase of the raw data in order to compare the results with that based on our own algorithm.

\section{Experimental Design and Data Analysis}\label{Exp:Sec}
In this section, the experiment is designed in order to obtain PRC of Drosophila (canton-S string) using ANF method as well as traditional double-actogram based method.

\subsection{Experimental Design}
During the experiment, the lab is securely locked to prevent any external stimulus (except the designated LED light) i.e, noise and ceiling light to affect the results. The experiment is designed such that all the flies are entrained to the same phase before the external light stimulus are applied:

\begin{itemize}
\item[(1)] Entrainment to the same phase: In order to investigate the influence of light stimulus to the circadian phase, we need to ensure that all the flies are entrained into the same phase before given the light stimulus. Therefore, a three days of LD 12:12 is imposed on all 9 incubators (each loaded with 21 flies).\\
\item[(2)] Light stimulus: After the three days entrainment, all incubators are put into darkness (DD). Light stimulus is programmed at the designated circadian time (4 lux blue light with duration of 1 hour) \footnote{The beginning of subjective night is defined as Circadian Phase (CP) 0.} (Table \ref{PRC:table}). \\
\end{itemize}

\begin{table}
\begin{center}
\begin{tabular}{|c|c|c|c|c|c|c|c|c|c|}
  \hline
  Light Stimulus (CP (hrs)): & N/A & 0 & 3 & 6 & 9 & 12 & 15 & 20 & N/A  \\\hline
  Light Strategy: & Dark (DD) & L & L & L & L & L & L & L & DD \\
  \hline

\end{tabular}
\end{center}
\caption{Blue light ($\lambda = 470$ nm) stimulus at the designated circadian time.\label{PRC:table}}
\end{table}

\medskip

\textbf{Data Analysis}: The raw data in the form of text files is transferred to PC which stores how many times each fly move along the tube within 1 minute with data logging time and light intensity information. There are more than 17280 data points for each channel (fly) with total 189 channels. The raw data is analyzed in the following manner:
\begin{itemize}
\item[(1)] Exclude the arrhythmic flies: We use periodogram analysis to detect the rhythmicity of a fly and examine the activity plot to exclude inactive flies. In our hands, we detect on average about $45\%$ flies (lowest incubator show approximately $30\%$) are rhythmic and will be used in the next step to obtain phase response curve.\\
\item[(2)] Utilize ANF to estimate the circadian phase: Each incubator's screened data are analyzed using ANF algorithm, and the estimated circadian phase is used to determine phase shift.\\
\item[(3)] Comparison to the traditional method using \emph{Cosinor} on double actogram to determine phase shift: The Acrophase (peak of a cosine wave fitted to the raw data) is used as a comparison in determining the phase shift (ActogramJ software \cite{ActoJ}). \\
\end{itemize}

\section{Results}\label{Result:Sec}
The experiments designed in this paper are standard, which have been widely adopted and discussed in the literature \cite{Klarsfeld03}. The intention of this paper is to demonstrate a new circadian phase estimation tool, ANF, that is accurate, continuous and robust against noise. We will first demonstrate the validity of our experiment, from which the PRC is obtained using ANF and compares with that results from literature as well as the traditional method based PRC (double-actogram and Acrophase); finally we demonstrate the filtering effect of ANF in which noise is added to the raw data and ANF successfully rejects the noise, while the traditional double-actogram method shows large error.

\subsection{Periodogram and the Exclusion of the arrhythmic flies}\label{SelectFly:Sec}

Raw activity data of each individual fly under the 3 days of LD entrainment and designated light stimulus (Table \ref{PRC:table}) are obtained and analyzed here in order to exclude arrhythmic and inactive flies.

\begin{figure}
        \centering
        \begin{subfigure}[t]{0.45\textwidth}
                \includegraphics[width=\textwidth]{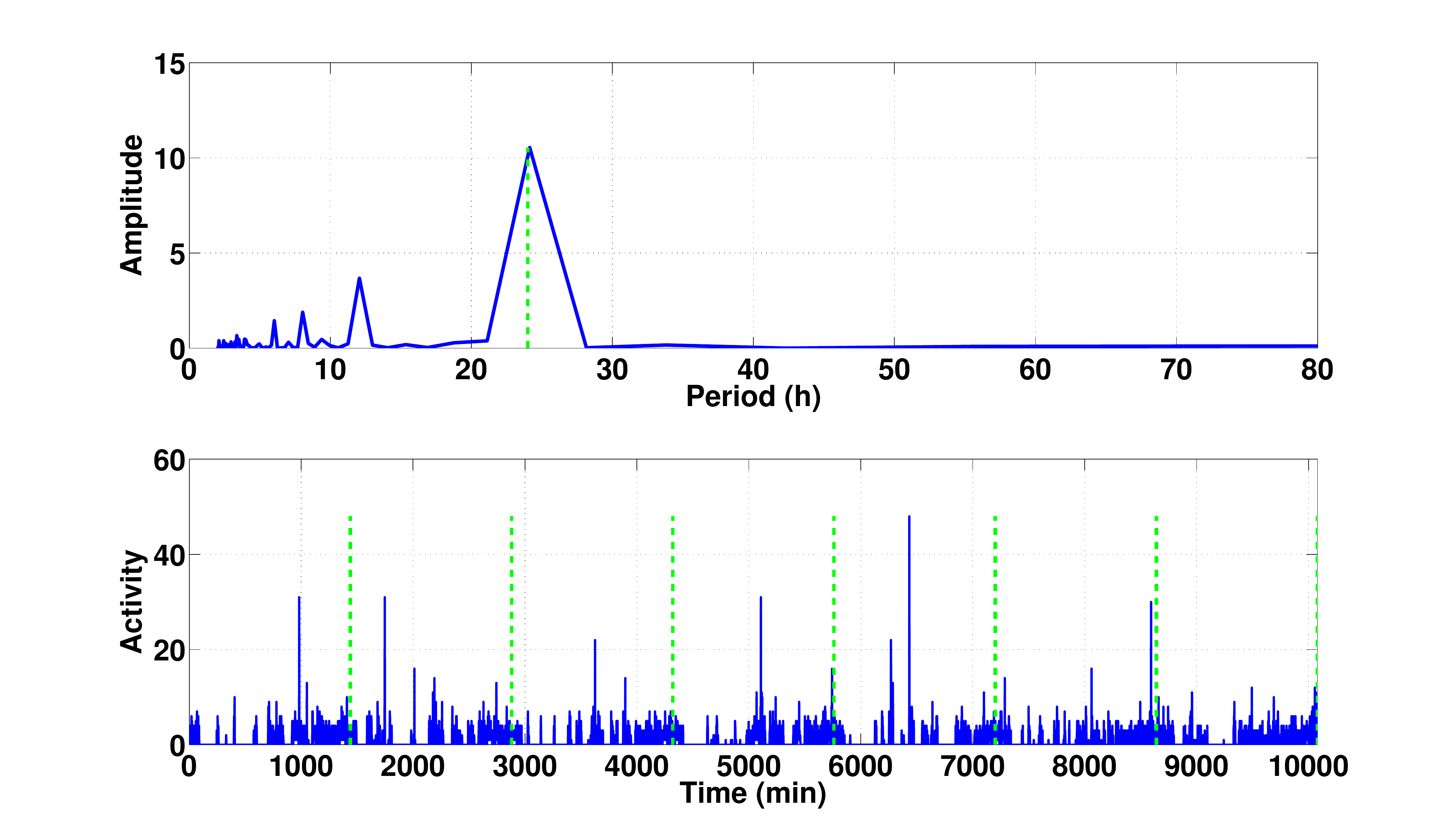}
                \caption{The fly passed the periodogram analysis which derived from activity records of individual flies in complete darkness. Top panel shows the periodogram and the bottom panel shows the raw activity records.\label{periodogram_24:fig}}
        \end{subfigure}
        \qquad
        \begin{subfigure}[t]{0.45\textwidth}
                \includegraphics[width=\textwidth]{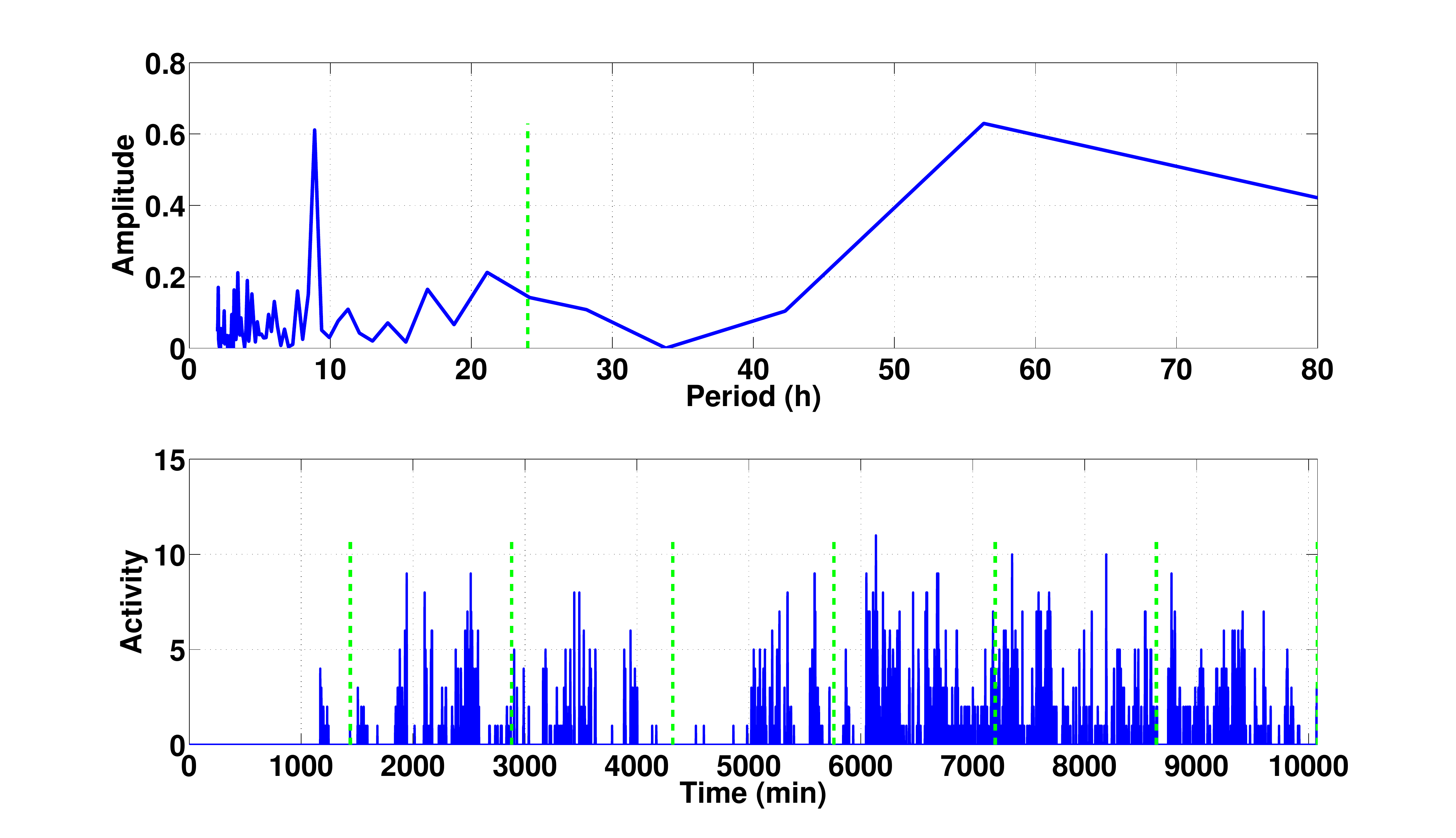}
                \caption{The fly failed the periodogram analysis which derived from activity records of individual flies in complete darkness. Top panel shows the periodogram and the bottom panel shows the raw activity records.\label{periodogram_bad:fig}}
        \end{subfigure}
        \caption{Periodogram analysis.}
\end{figure}

Periodogram is first applied to exclude the flies with an irregular period. \footnote{Periodogram is a nonparametric estimate of the power spectral density (PSD) estimate of the signal and uses traditional Fourier techniques such as fast Fourier transform (FFT).} The data sets analyzed by the periodogram are the activity records of individual flies from day 6 to day 12 when the flies are in complete darkness. An example is shown in Figure \ref{periodogram_24:fig} and Figure \ref{periodogram_bad:fig} where flies with regular period and irregular period are shown, respectively. The averaged inclusion ratio is about $45\%$ in each incubator.

\subsection{Comparison of ANF and Acrophase based Phase Estimation}\label{PRCacto:section}
\begin{figure}
        \centering
        \begin{subfigure}[t]{0.45\textwidth}
                \includegraphics[width=\textwidth]{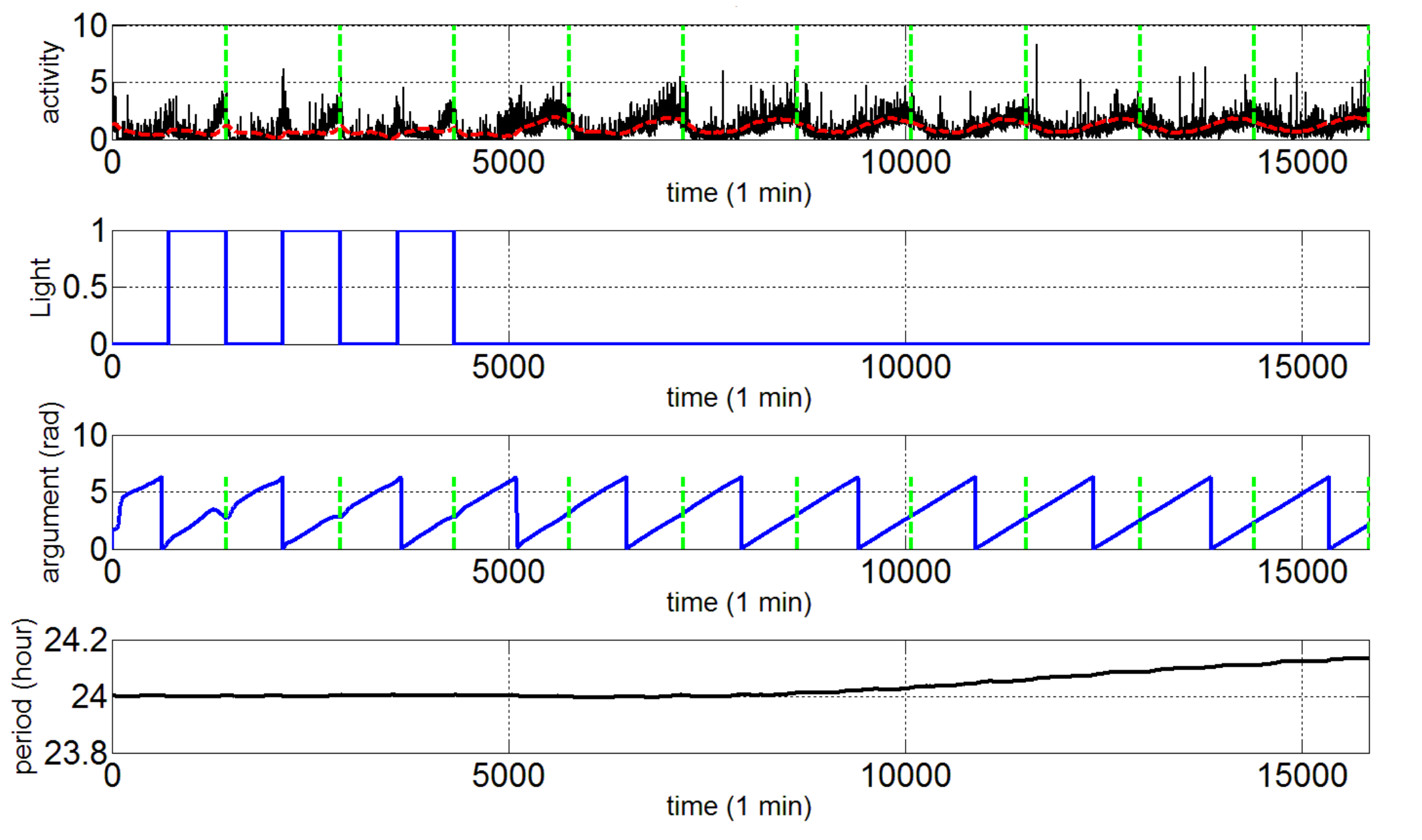}
                \caption{Incubator with the 3 days of LD 12:12 and then free run. \label{Monitor2:fig}}
        \end{subfigure}
        \qquad
        \begin{subfigure}[t]{0.45\textwidth}
                \includegraphics[width=\textwidth]{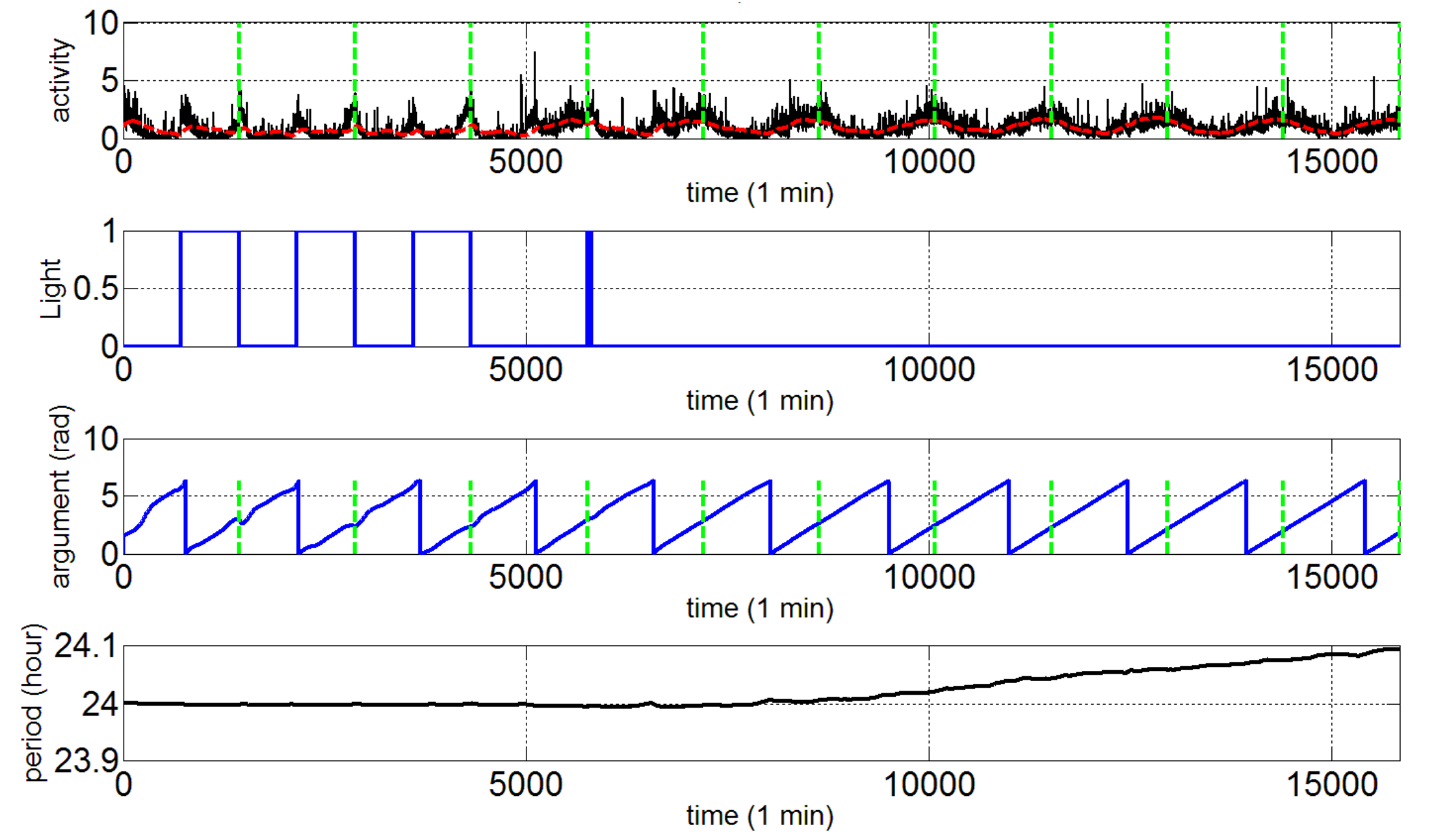}
                \caption{Incubator with the 3 days of LD 12:12 and then a light pulse at CP 0 with 4 lux intensity and 1 h duration.\label{Monitor4:fig}}
        \end{subfigure}
        \caption{ANF analysis of the averaged data of each incubator: top panel shows raw activity (bold black) with standard deviation (light black) and the ANF estimation (red); $2{nd}$ panel shows the light pattern; bottom panel shows the ANF period adaptation.}
\end{figure}

By excluding the arrhythmic flies, we use the selected data set to obtain phase response curve. The raw activity data from the selected flies of each incubator are first averaged and then analyzed by the ANF algorithm (Section \ref{ANF:Sec}) to determine the phase of the averaged flies in each incubator. Results of two incubators are shown in Figure \ref{Monitor2:fig} and Figure \ref{Monitor4:fig} in which 3 days of LD entrainment followed by a free running and a light pulse at CP 0, respectively. Top panel shows raw activity (black) and the ANF estimation (red); $2{nd}$ panel shows the light pattern; bottom panel shows the ANF estimation of circadian phase. Notice that ANF provides a continuous period adaptation and circadian phase estimation, which will be a very useful tool in any circadian related research. Now the only question remains is how accurate the ANF based method is. We will compare the ANF based method with the traditional double-actogram based method using the same data set.

\begin{figure}
        \centering
        \begin{subfigure}[t]{0.3\textwidth}
                \includegraphics[width=\textwidth]{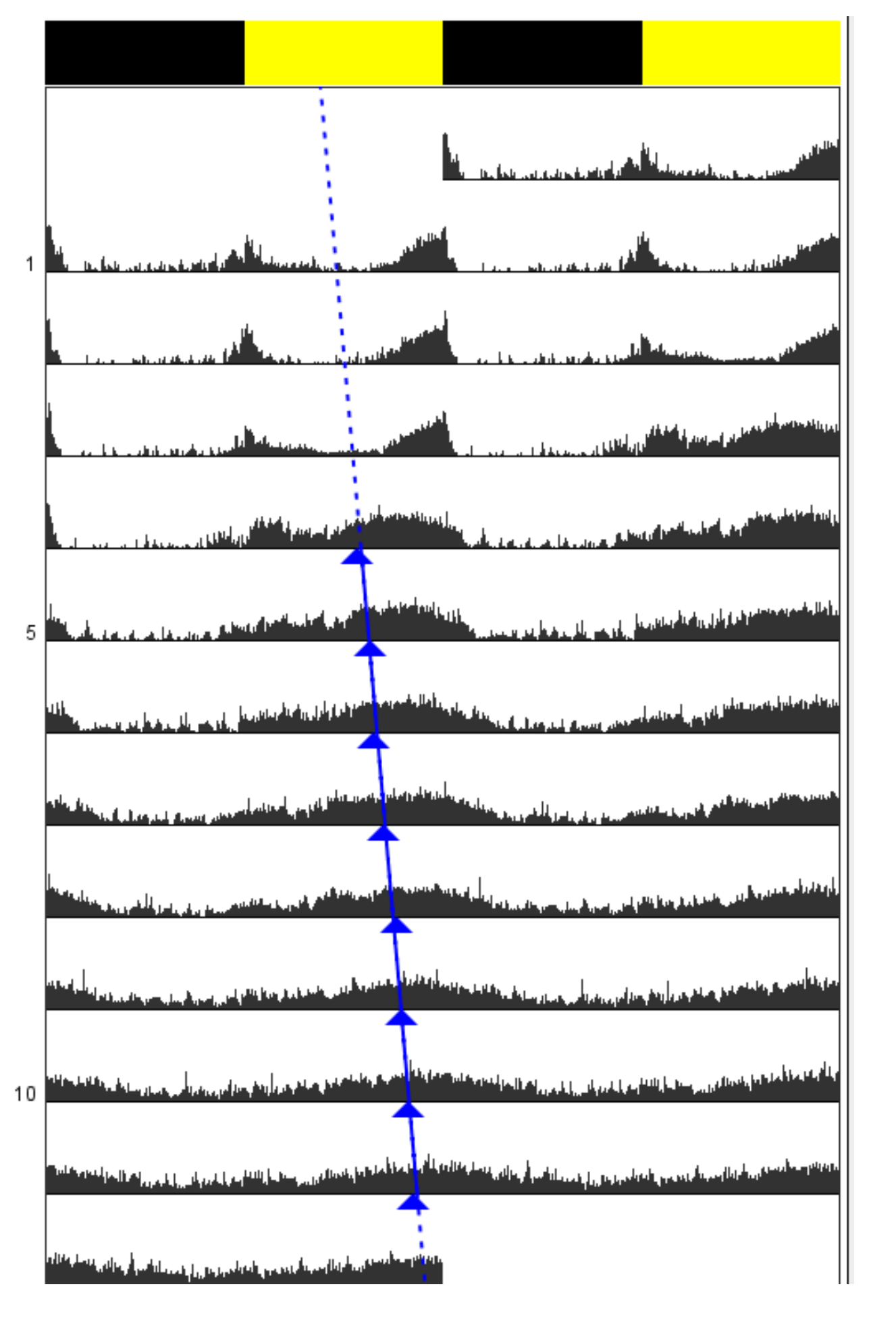}
                \caption{Double actogram and the Acrophase estimation using ActogramJ. Blue triangles are the estimated daily Acrophase and the blue lines is a regression fit. \label{ActJ_Mo2_Act:fig}}
        \end{subfigure}
        \qquad
        \begin{subfigure}[t]{0.5\textwidth}
                \includegraphics[width=\textwidth]{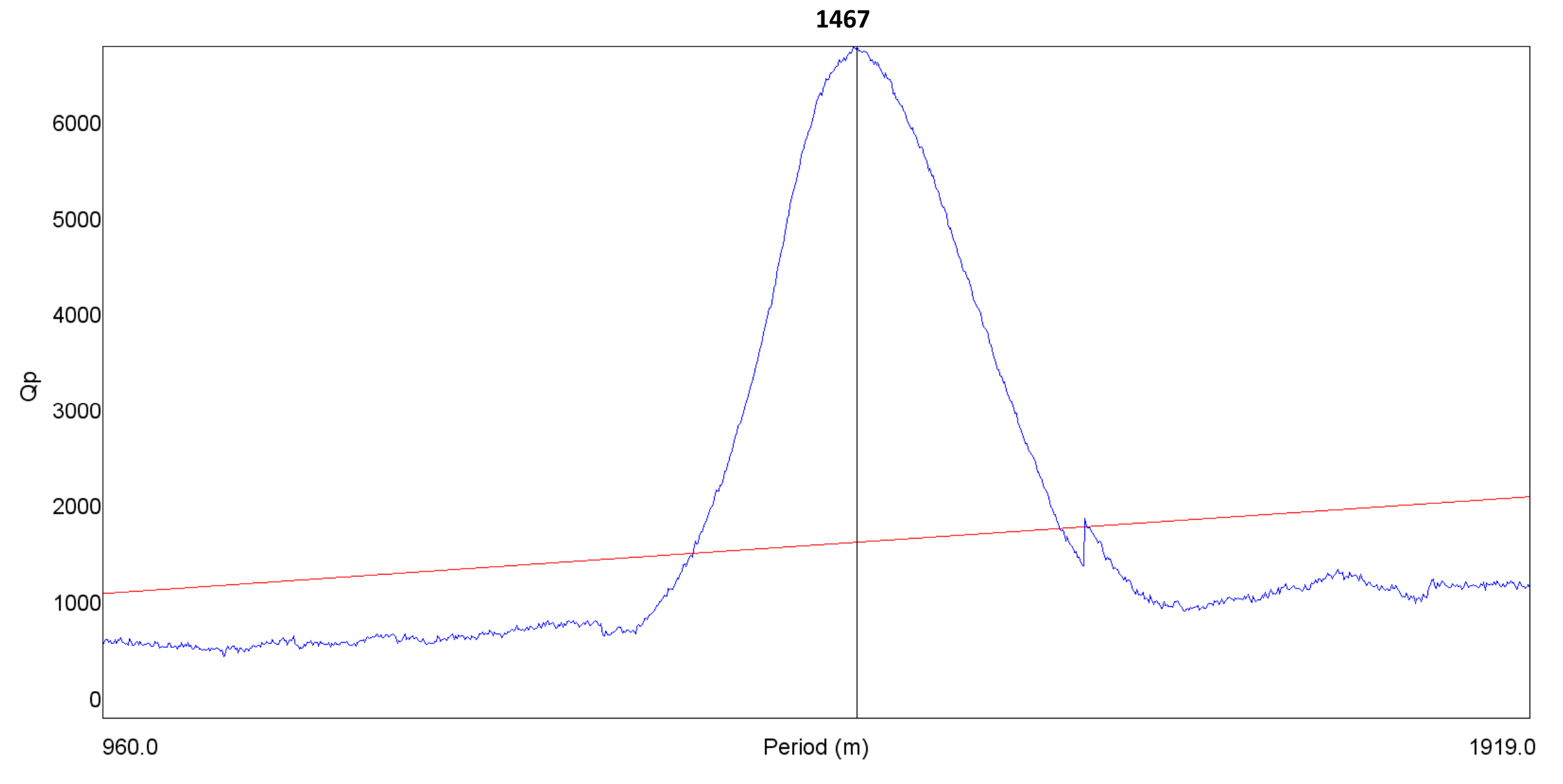}
                \caption{Periodogram after the last light off using ActogramJ.\label{ActJ_Mo2_Per:fig}}
        \end{subfigure}
        \caption{Obtaining Periodogram and Acrophase based on the double actogram using ActogramJ (3 days of LD 12:12 and then free run).}
\end{figure}

The traditional method in determining circadian phase is mostly based on double-actogram where the daily peak of the activity is identified using eye fit or Acrophase (peak of a cosine fitted curve). Here we use ActogramJ software \cite{ActoJ} to obtain Acrophase based on the same raw activity data of the selected flies obtained from Section \ref{SelectFly:Sec}. The double-actogram of control incubator is shown in Figure \ref{ActJ_Mo2_Act:fig} where blue triangles are the estimated daily Acrophase (with a regression fitted line) and the periodogram from day 4-12 is shown in Figure \ref{ActJ_Mo2_Per:fig}. Notice here the free-running period is $\tau = 24.45$\footnote{Period of another control incubator is $\tau = 24.3$.} where the ANF based period estimation is trying to adapt shown in Figure \ref{Monitor2:fig} and \ref{Monitor4:fig} bottom panel.

With the analysis of both methods ready, we move on to obtaining phase response curve. Note that circadian phase is a relative concept, especially when obtaining PRC (the free-running period is normally not exactly 24 hrs). Therefore the control incubator with 3 days of LD entrainment followed by free-running is used as the reference incubator (Figure \ref{Monitor2:fig}) and agrees with the standard analysis in the literature. Phase shift is then defined as the phase difference between the target incubator and the control incubator. The same analysis is applied to both ANF and double-actogram (ActogramJ) based method.

\begin{figure}[ht]
\bc
  \includegraphics[width=0.4\textwidth]{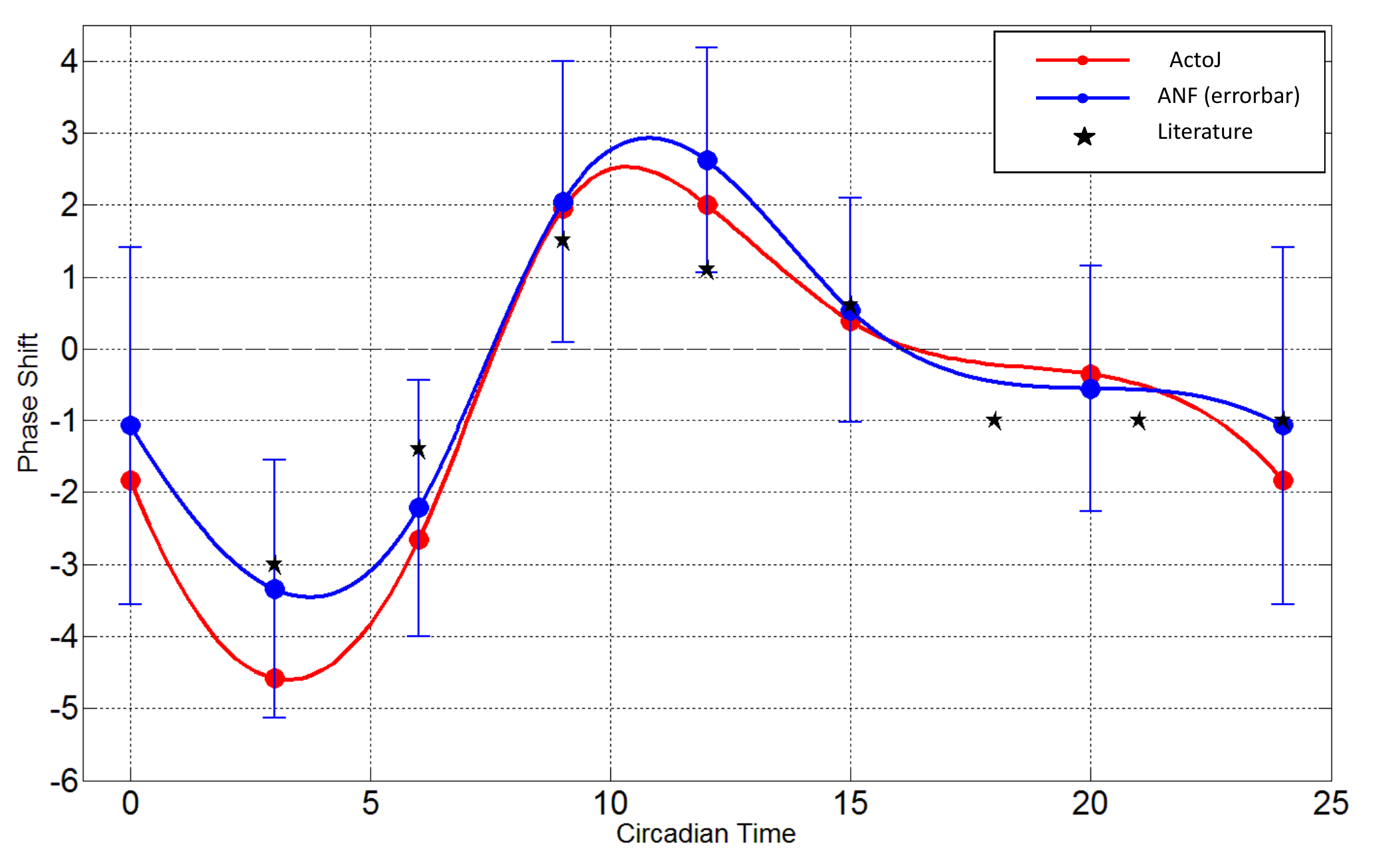}
\ec
  \caption{ActogramJ based PRC (red dots and interpolated red curve) and the PRC obtained by ANF (blue dots and interpolated blue curve), PRC obtained in the \cite{Klarsfeld03} is shown as black stars.\label{PRC_actJ:fig}}
\end{figure}

Two PRCs of both methods are shown in Figure \ref{PRC_actJ:fig}\footnote{Standard deviation of the ANF based method on individual fly in each incubator is shown as error bar, which is calculated as the square root of the sum of the variance of target and control incubator.} based on the phase difference between target incubator and control incubator on day 10 with light stimulus given in table \ref{PRC:table}. Notice the results are almost identical, demonstrating that the accuracy of the ANF based method meets the well-accepted standard. We also compare our results with those obtained in the literature \cite{Klarsfeld03} (black stars in Figure \ref{PRC_actJ:fig}). The ANF obtains very similar results compared to what has been accepted by the literature.

\subsection{The Filtering Effect of ANF}\label{PRCnoise:section}
The validity of ANF based circadian phase estimation is demonstrated above, where the result is almost identical to that obtained by double-actogram based method and matches the PRC obtained by other researchers very well \cite{Klarsfeld03}. In other words, ANF not only meets the accepted standard in the literature but also provides a continuous circadian phase estimation.

\begin{figure}
        \centering
        \begin{subfigure}[t]{0.45\textwidth}
                \includegraphics[width=\textwidth]{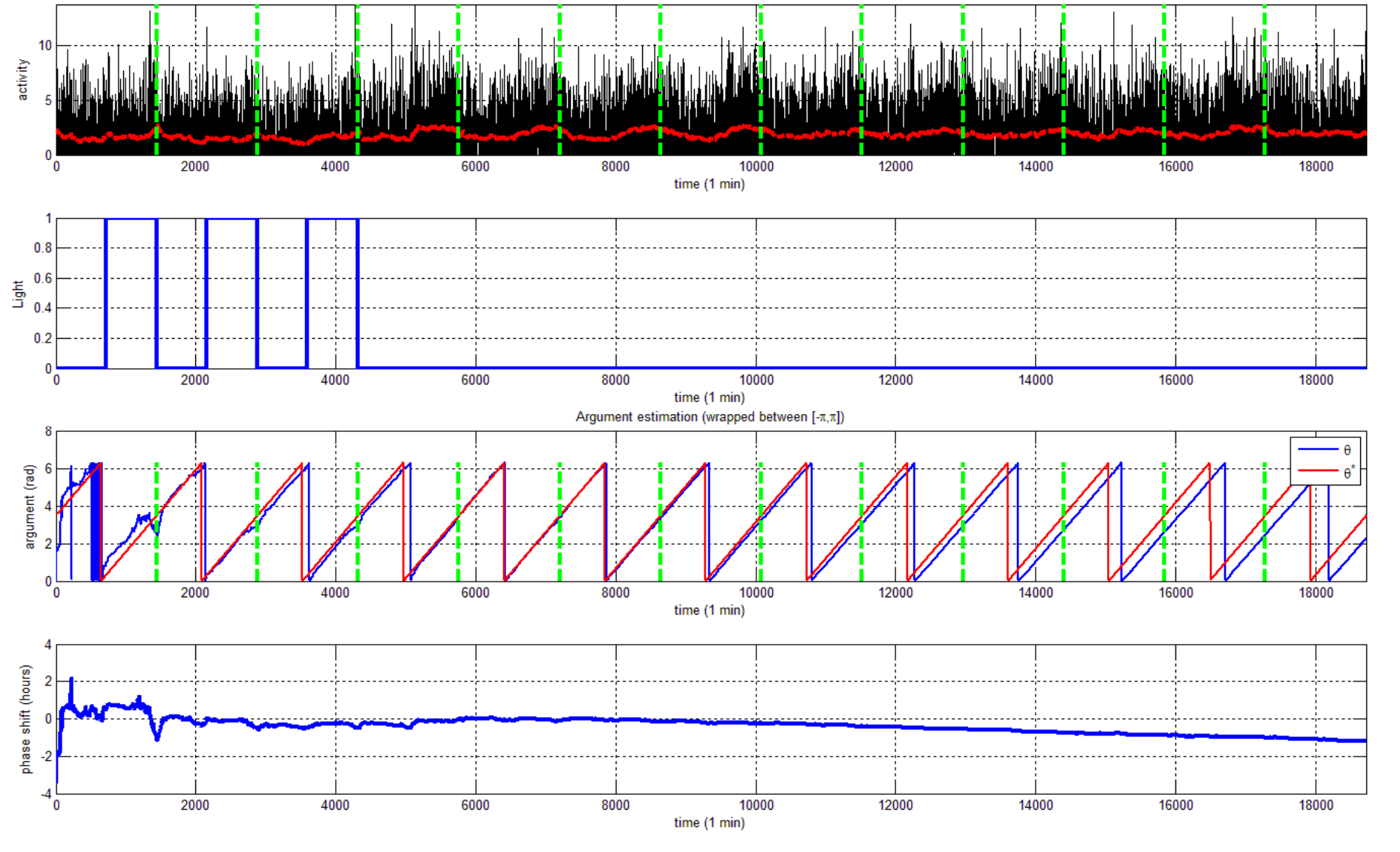}
                \caption{Incubator with the 3 days of LD 12:12 and then free run. \label{Cor_Monitor2:fig}}
        \end{subfigure}
        \qquad
        \begin{subfigure}[t]{0.45\textwidth}
                \includegraphics[width=\textwidth]{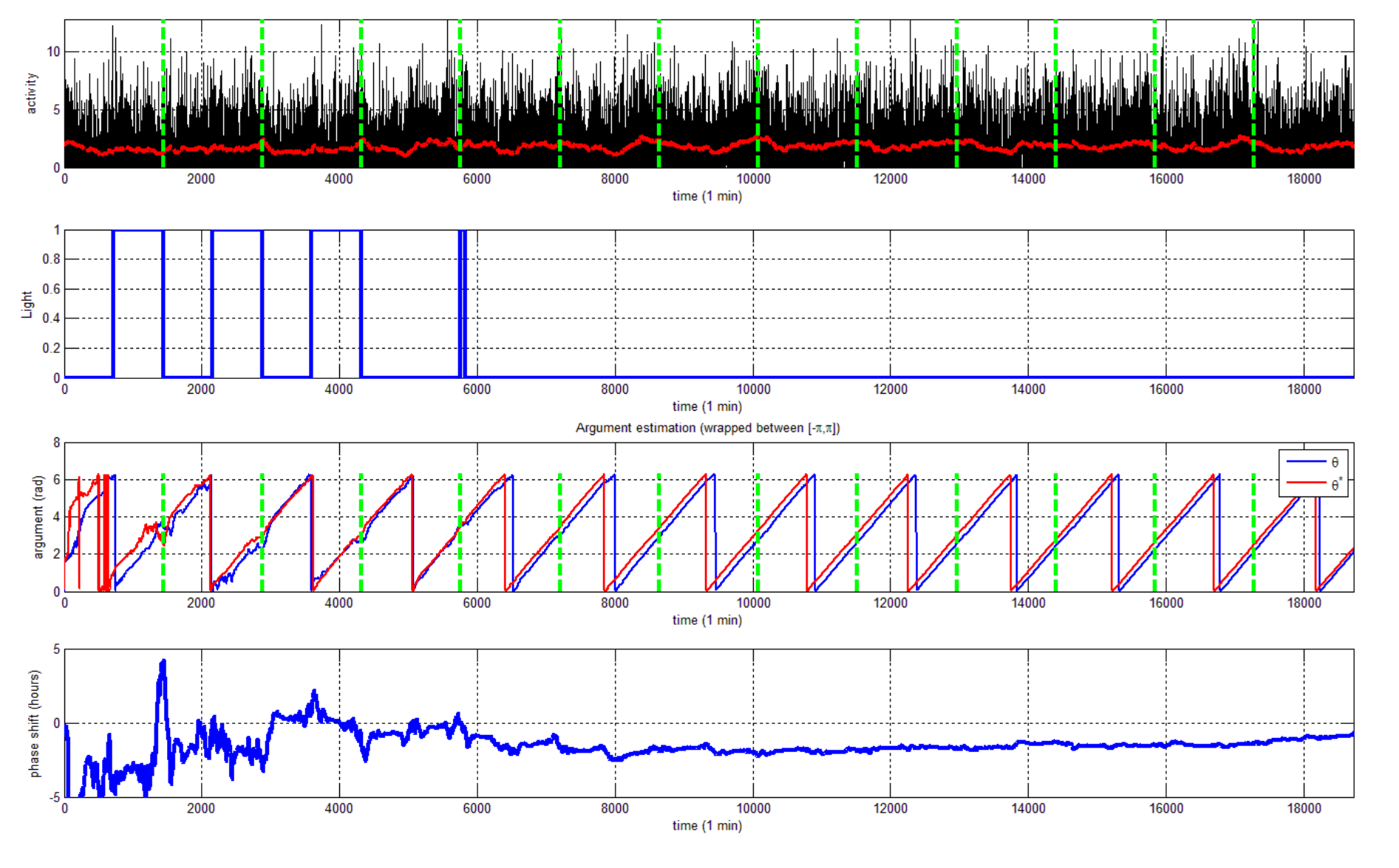}
                \caption{Incubator with the 3 days of LD 12:12 and then a light pulse at CP 0 with 4 lux intensity and 1 h duration.\label{Cor_Monitor4:fig}}
        \end{subfigure}
        \caption{ANF analysis of the corrupted data with noise $w$ of each incubator: top panel shows raw activity (bold black) with standard deviation (light black) and the ANF estimation (red); $2{nd}$ panel shows the light pattern; bottom panel shows the ANF estimation of circadian phase.}
\end{figure}

\begin{figure}[ht]
\bc
  \includegraphics[width=0.4\textwidth]{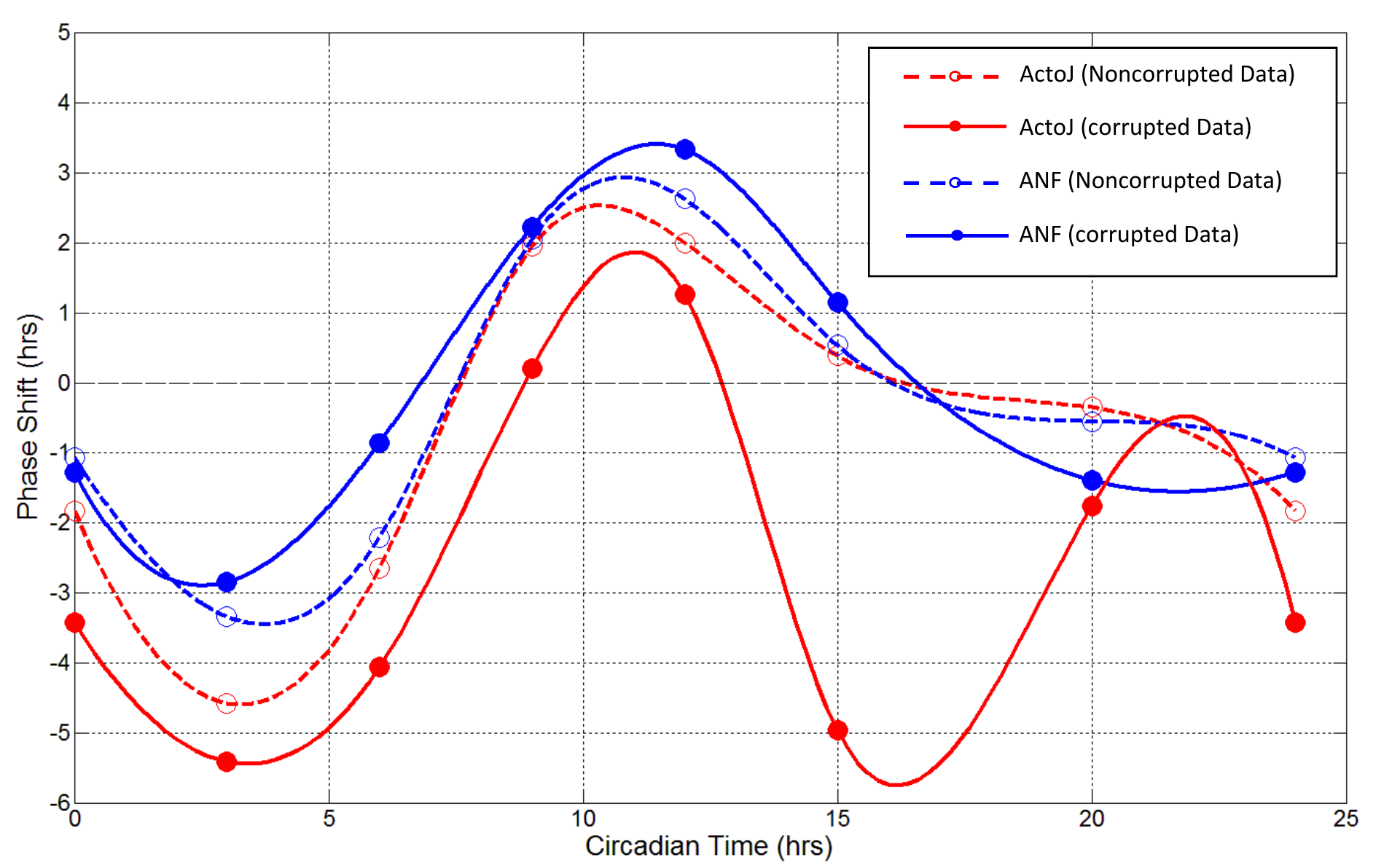}
\ec
  \caption{PRC obtained through the corrupted data with noise $w$, by ANF (blue dots and interpolated blue curve) and ActogramJ (red dots and interpolated red curve). The original (based on non-corrupted data) PRCs are shown as dotted curves.\label{Cor_PRC:fig}}
\end{figure}

Yet another important feature of ANF is the robustness against higher order harmonics, or noise \cite{ZhangACC13}. In order to demonstrate this feature, we intentionally corrupt the raw activity data selected in Section \ref{SelectFly:Sec} with zero-mean Gaussian white noise $w$ where variance $Var(w) = 10$. Two examples are shown in Figure \ref{Cor_Monitor2:fig} and \ref{Cor_Monitor4:fig}. The same procedure is applied to obtain ANF based PRC and double-actogram based PRC (ActoJgramJ) and the result is shown in Figure \ref{Cor_PRC:fig}. Notice that the ANF based method successfully rejects the noise effect and the error against its original PRC is very small (average error is 0.57 hrs), while the double-actogram based method shows large error (average error is 1.84 hrs).

\section{Conclusion}\label{Conc:Sec}

\section*{Acknowledgment}
This work was supported primarily by the Army Research Office through Grant number W911NF-13-1-0265. We also gratefully acknowledge the support from National Science Foundation (NSF) through the Smart Lighting Engineering Research Center (EEC-0812056) and from the Center for Automation Technologies and Systems (CATS), which  under a block grant from the New York State Empire State Development Division of Science, Technology and Innovation (NYSTAR) contract C090145. The authors would also like to thank Drs. Mark Rea, Mariana Figueiro, and Andrew Bierman at the RPI Light Research Center for introducing this topic to the authors.

\bibliography{Wei_circadian_ref}

\begin{thebibliography}{10}

\bibitem{kripke1978circadian}
D.F. Kripke, D.J. Mullaney, M.~Atkinson, and S.~Wolf.
\newblock Circadian rhythm disorders in manic-depressives.
\newblock {\em Biological Psychiatry}, 1978.

\bibitem{knutsson03}
A.~Knutsson.
\newblock Health disorders of shift workers.
\newblock {\em Occupational Medicine}, 53(2):103--108, 2003.

\bibitem{sephton2003circadian}
S.~Sephton and D.~Spiegel.
\newblock Circadian disruption in cancer: A neuroendocrine-immune pathway from
  stress to disease?
\newblock {\em Brain, Behavior, and Immunity}, 17(5):321--328, 2003.

\bibitem{stevens2005circadian}
R.G. Stevens.
\newblock Circadian disruption and breast cancer: from melatonin to clock
  genes.
\newblock {\em Epidemiology}, 16(2):254, 2005.

\bibitem{Rea08}
M.S Rea, A.~Bierman, M.G. Figueiro, and J.D. Bullough.
\newblock A new approach to understanding the impact of circadian disruption on
  human health.
\newblock {\em Journal of Circadian Rhythm}, 6, 2008.

\bibitem{kelly1999nonentrained}
T.L. Kelly, D.F. Neri, J.T. Grill, D.~Ryman, P.D. Hunt, D.J. Dijk, T.L.
  Shanahan, and C.A. Czeisler.
\newblock Nonentrained circadian rhythms of melatonin in submariners scheduled
  to an 18-hour day.
\newblock {\em Journal of Biological Rhythms}, 14(3):190, 1999.

\bibitem{mills1964circadian}
J.N. Mills.
\newblock Circadian rhythms during and after three months in solitude
  underground.
\newblock {\em The Journal of Physiology}, 174(2):217, 1964.

\bibitem{harrington10}
M.~Harrington.
\newblock Location, location, location: important for jet-lagged circadian
  loops.
\newblock {\em Journal of Clinical Investigation}, 120(7):2265--2267, July
  2010.

\bibitem{Leloup99}
J.C. Leloup, D.~Gonze, and A.~Goldbeter.
\newblock Limit cycle models for circadian rhythms based on transcriptional
  regulation in drosophila and neurospora.
\newblock {\em J. Biological Rhythms}, 14(6):433--448, 1999.

\bibitem{ZhangACC13}
J.~Zhang, J.T. Wen, and A.~Julius.
\newblock Adaptive circadian rhythm estimator and its application to circadian
  rhythm control.
\newblock In {\em American Control Conference}, Washington D.C, June 2013.

\bibitem{Hsu1999}
L.~Hsu, R.~Ortega, and G.~Damm.
\newblock A globally convergent frequency estimator.
\newblock {\em IEEE Trans. Automat. Control}, 44(4):698--713, 1999.

\bibitem{Frank69}
K.D. Frank and W.F. Zimmerman.
\newblock Action spectra for phase shifts of a circadian rhythm in drosophila.
\newblock {\em Science}, 163(3868):688--689, 1969.

\bibitem{Klarsfeld03}
A.~Klarsfeld, J.C. Leloup, and F.~Rouyer.
\newblock Circadian rhythms of locomotor activity in drosophila.
\newblock {\em Behavioural Processes}, 64:161--175, 2003.

\bibitem{Emery00}
P.~Emery, R.~Stanewsky, and C.H. Forster.
\newblock Drosophila cry is a deep brain circadian photoreceptor.
\newblock {\em Neuron}, 26:493¨C504, 2000.

\bibitem{ActoJ}
B.~Schmid, C.~Helfrich-Forster, and T.~Yoshii.
\newblock A new imagej plug-in ''actogramj'' for chronobiological analysis.
\newblock {\em Journal of Biological Rhythms}, 26(5):464--467, 2011.

\end{thebibliography}
\end{document}